\documentclass[aps,showpacs,twocolumn,superscriptaddress,prl]{revtex4}
\usepackage[pdftex]{graphicx}
\usepackage{times,xspace}
\usepackage{amsbsy,amssymb,amsmath,bm}
\usepackage{dcolumn}% Align table columns on decimal point
\usepackage{bm}% bold math
\input{epsf.sty}

\def\bbbc{{\mathchoice {\setbox0=\hbox{$\displaystyle\rm C$}\hbox{\hbox
to0pt{\kern0.4\wd0\vrule height0.9\ht0\hss}\box0}}
{\setbox0=\hbox{$\textstyle\rm C$}\hbox{\hbox
to0pt{\kern0.4\wd0\vrule height0.9\ht0\hss}\box0}}
{\setbox0=\hbox{$\scriptstyle\rm C$}\hbox{\hbox
to0pt{\kern0.4\wd0\vrule height0.9\ht0\hss}\box0}}
{\setbox0=\hbox{$\scriptscriptstyle\rm C$}\hbox{\hbox
to0pt{\kern0.4\wd0\vrule height0.9\ht0\hss}\box0}}}}

\begin{document}
\title{Thermodynamic and transport properties of   NiCl$_2$-4SC(NH$_2$)$_2$: role of strong mass renormalization}

%\author{Y. Kohama$^1$, N. Dilley$^2$, A. V. Sologubenko$^3$, V.S. Zapf$^1$, M. Jaime$^1$, J. Mydosh$^4$, A. Paduan-Filho$^5$, K. Al-Hassanieh$^6$, P. Sengupta$^7$, S. Gangadharaiah$^8$, A. L. Chernyshev$^8$, C.D. Batista$^6$}

%\affiliation{

%$^1$MPA-CMMS, Condensed Matter and %Material Sciences group, LANL \\
%$^3$II.Physikalisches Institut, Universit\{a}t zu K\{o}ln, Z\{u}lpicher Str. 77, 50937 K\{o}ln, Germany \\
%$^2$Quantum Design Inc., San Diego, CA \\
%$^4$Kamerlingh Onnes Laboratory, Leiden University, 2300RA Leiden, The Netherlands \\
%$^5$Instituto de Fisica, Universidade de Sao Paulo, Brazil \\}
%$^6$T-4, Los Alamos National Lab (LANL), Los Alamos, NM \\
%$^7$School of Physical and Mathematical Sciences, Nanyang Technological University, 50 Naynag Avenue, Singapore 639798 \\
%$^8$ Department of Physics and Astronomy, University of California, Irvine, CA  92697, USA\\
%}

\author{Y. Kohama}
\affiliation{MPA-CMMS, LANL, Los Alamos, New Mexico 87545, USA}

\author{A. V. Sologubenko} 
\affiliation{{\protect II.}\ Physikalisches Institut, Universit\"at zu K\"oln, Z\"ulpicher Str.\ 77, 50937 K\"oln, Germany}
%\affiliation{II.Physikalisches Institut, Universitat zu Koln, Zlpicher Str. 77, 50937 Koln, Germany }

\author{N. R. Dilley}
\affiliation{Quantum Design Inc., San Diego, CA 92121, USA}

\author{V. S. Zapf}
\affiliation{MPA-CMMS, LANL, Los Alamos, New Mexico 87545, USA}

\author{M. Jaime}
\affiliation{MPA-CMMS, LANL, Los Alamos, New Mexico 87545, USA}

\author{J. Mydosh} 
\affiliation{Kamerlingh Onnes Laboratory, Leiden University, 2300RA Leiden, The Netherlands}

\author{A. Paduan-Filho} 
\affiliation{Instituto de Fisica, Universidade de Sao Paulo, Brazil}

\author{K. Al-Hassanieh}
\affiliation{Theoretical Division, Los Alamos National Laboratory, Los Alamos, New Mexico 87545, USA}

\author{P. Sengupta}
\affiliation{School of Physical and Mathematical Sciences, Nanyang Technological University, 50 Naynag Avenue, Singapore 639798}

\author{S. Gangadharaiah}
\affiliation{Department of Physics and Astronomy, University of California, Irvine, CA  92697, USA}
\affiliation{Max-Planck-Institut f\"ur Physik komplexer Systeme, N\"othnitzer Str.\ 38,  01187 Dresden, Germany}

\author{A. L. Chernyshev} 
\affiliation{Department of Physics and Astronomy, University of California, Irvine, CA  92697, USA}
\affiliation{Max-Planck-Institut f\"ur Physik komplexer Systeme, N\"othnitzer Str.\ 38,  01187 Dresden, Germany}

\author{C. D. Batista}
\affiliation{Theoretical Division, Los Alamos National Laboratory, Los Alamos, New Mexico 87545, USA}

\date{\today}

\begin{abstract}

Several quantum paramagnets exhibit magnetic field-induced  quantum phase transitions to an antiferromagnetic state that exists for $H_{c1}\!\leq\!H\!\leq\! H_{c2}$.  For some of these compounds, there is a significant asymmetry between the low- and high-field transitions. We present specific heat and thermal conductivity measurements in NiCl$_2$-4SC(NH$_2$)$_2$, together with calculations which show that the asymmetry is caused by a strong  mass renormalization due to quantum fluctuations for $H\!\leq\!H_{c1}$ that are absent for $H\!\geq\!H_{c2}$. We argue that the enigmatic lack of asymmetry in thermal conductivity is due to a concomitant renormalization of the impurity scattering.

\end{abstract}

%---------------------------------------------------------------

\pacs{75.10.Jm,   % Quantized spin models
      75.40-s,   % 
      75.40.Cx    %
      }

%---------------------------------------------------------------
\maketitle

The correspondence between a spin system and a gas of bosons has been very fruitful for describing field-induced ordered phases in a large class of 
quantum paramagnets \cite{Affleck91,Thierry99,Nikuni00,Jaime04,Sebastian06b}. In this analogy, a magnetic field $H$ plays the role of the 
chemical potential, which, upon reaching a critical value $H_{c1}$, induces a $T\!=\!0$ Bose-Einstein condensation (BEC), provided that the number 
of bosons is conserved, the kinetic energy is dominant, and the spatial dimension $d\!>\!1$. Such a BEC state corresponds to a canted XY magnetic 
ordering of the spins. 

At the BEC quantum critical point (QCP), the low-energy bosonic excitations have a quadratic 
dispersion $\omega\!=\!k^2/2m^*$, where $m^*$ is the effective mass. This mass is renormalized by quantum fluctuations  in the paramagnetic phase $H\!\leq\!H_{c1}$ . In magnets with $H_{c1}\!\ll\!H_{c2}$ the renormalization can be expected to be very strong because of the 
proximity to the magnetic instability. The transition at $H_{c1}$ should be contrasted with the second BEC-QCP that takes place at the saturation 
field $H_{c2}$ \cite{Batyev84}. Since the field induced magnetization is a conserved quantity, there are no quantum fluctuations and no
mass renormalization for the fully polarized phase above $H_{c2}$, i.e., the bare mass $m$  can be obtained from the single-particle excitation spectrum at $H\!\geq\!H_{c2}$. Thus, quantum paramagnets are ideal for studying mass renormalization effects because the effective and the bare bosonic masses can be obtained from 
{\em two different QCP's that occur in the same material}.

Here we present theoretical and experimental evidence for a strong mass renormalization effect, $m/m^*\simeq 3$, in 
%the quantum magnet 
NiCl$_2$-4SC(NH$_2$)$_2$ [referred to as DTN]. We will show that  the large asymmetry between the peaks in the low-temperature specific heat, $C_v(H)$, in the vicinity of $H_{c1}$ and $H_{c2}$ is 
closely described by analytical and  Quantum Monte Carlo (QMC) calculations. The mass renormalization also explains 
similar asymmetries observed in  
other properties of DTN, such as  magnetization \cite{PaduanFilho04}, electron spin resonance \cite{Zvyagin07}, sound velocity 
\cite{Chiatti09,Zherlitsyn09}, and magnetostriction \cite{Zapf08}. In  a remarkable contrast to these properties, peaks in
the low-temperature thermal conductivity, $\kappa$, near $H_{c1}$ and $H_{c2}$ {\it do not} show any substantial asymmetry. We provide an explanation to this dichotomy by demonstrating that the leading boson-impurity scattering amplitude is also renormalized by quantum fluctuations, effectively canceling mass renormalization effect in $\kappa$. 

DTN is a quantum magnet with tetragonal crystal symmetry that exhibits a field-induced BEC \cite{PaduanFilho04,Zapf06,Zvyagin07,Zvyagin08,Yin08} 
to a very good approximation \cite{note}. The dominant single-ion uniaxial anisotropy $D\!=\!8.9$~K splits the Ni $S\!=\!1$ triplet into 
 an $S^z\!=\!0$ ground state and  an $S^z\!=\!\pm 1$ excited doublet. The antiferromagnetic exchange coupling between Ni ions 
is $J_c\!=\!2.2$~K along the 
\textsl{c}-axis and $J_a\!=\!0.18$~K along the \textsl{a}- and \textsl{b}-axes, while the gyromagnetic factor along the 
$c$-axis is $g=2.26$ \cite{Zvyagin07}. A magnetic field applied along the \textsl{c}-axis lowers the energy of the $S^z\!=\!1$ 
state producing a 
$T\!=\!0$ BEC transition at $H_{c1}\!=\!2.1$~T. The long-range order occurs in a dome-shaped region of the $T\!-\!H$ phase diagram between 
$H_{c1}$ and  $H_{c2}\!=\!12.5$~T and below the maximum ordering temperature  $T_{max}\!\simeq\!1.2$K~\cite{Zapf06}. The $T^{3/2}$  dependence of the critical field expected for a BEC-QCP has been established via direct measurements of the phase boundary with ac susceptibility 
down to 1~mK \cite{Yin08}, and by magnetization measurements \cite{PaduanFilho09}. The asymmetry between $H_{c1}$ and $H_{c2}$ 
\cite{Zapf06,PaduanFilho04,Zvyagin07,Chiatti09,Zherlitsyn09,Zapf08}  can also be seen directly in the skewed shape of the phase diagram \cite{Zapf06}.

The Hamiltonian describing the $S\!=\!1$ spin degrees of freedom of DTN in external field is given by \cite{Zvyagin07} 
\begin{eqnarray}
{\cal H} = \sum_{{\bf r},\nu} J_{\nu}\, {\bf S}_{\bf r} \cdot {\bf S}_{{\bf r}+{\bf e}_{\nu}} 
+D \sum_{{\bf r}} \left(S^z_{{\bf r}}\right)^2 - h\sum_{{\bf r}} S^z_{{\bf r}},
\label{H1}
\end{eqnarray}
where ${\bf e}_{\nu}$ are the primitive vectors of the lattice, $\nu\!=\!\{a, b, c\}$, and $h\!=\!g \mu_B H$. We introduce  
Schwinger bosons associated with the fundamental representation of SU(3) that obey the constraint
$\sum_m b^{\dagger}_{{\bf r}m} b^{\;}_{{\bf r}m}\!=\!1$. The subscript $m\!=\!\{\downarrow,0,\uparrow \}$ labels the eigenstates of $S^z_{\bf r}$ 
with the eigenvalues $\{-1,0,1\}$.  The spin operators in this representation are:
\begin{eqnarray}
S^{z}_{\bf r} =   n_{{\bf r}\uparrow} - n_{{\bf r}\downarrow}, \;
%b^{\dagger}_{{\bf r}\uparrow} b^{\;}_{{\bf r}\uparrow} - b^{\dagger}_{{\bf
%r}\downarrow} b^{\;}_{{\bf r}\downarrow} 
%\nonumber \\
S^{+}_{\bf r} =  \left(S^{-}_{\bf r} \right)^\dag=
\sqrt{2} \,\left(b^{\dagger}_{{\bf r}\uparrow} b^{\;}_{{\bf r}0} + 
b^{\dagger}_{{\bf r}0} b^{\;}_{{\bf r}\downarrow}\right), \, 
\label{sop}
\end{eqnarray}
with $n_{{\bf r}m} = b^{\dagger}_{{\bf r}m} b^{\;}_{{\bf r}m}$.
We enforce the constraint by introducing spatially uniform Lagrange multiplier $\mu$ 
\begin{eqnarray}
{\hat{\cal H}} = {\cal H} + \mu\sum_{\bf r} \left(
b^{\dagger}_{{\bf r}\uparrow} b^{\dagger}_{{\bf r}\uparrow} + 
b^{\dagger}_{{\bf r}\downarrow} b^{\dagger}_{{\bf r}\downarrow}
+ b^{\dagger}_{{\bf r}0} b^{\dagger}_{{\bf r}0} -1 \right). 
\end{eqnarray}
The lowest energy state in the $H\!<\!H_{c1}$ paramagnetic regime is $b^{\dagger}_{{\bf r}0} |0\rangle$  and the ground state corresponds to a non-zero expectation value  of the $S^z\!=\!0$ boson: $b^{\dagger}_{{\bf r}0}\!=\!b^{\;}_{{\bf r}0}\!=\!s$. By using the spin representation 
\eqref{sop} with the mean-field value for $b^{(\dagger)}_{0}$ and neglecting higher-order terms in powers of $b^{(\dagger)}_{\uparrow(\downarrow)}$, 
we obtain the Hamiltonian in the harmonic approximation
\begin{eqnarray}
{\hat{\cal H}}= E_0 +\sum_{{\bf k},\sigma} \Big[
A_{{\bf k}{ \sigma}}
{\hat b}^{\dagger}_{{\bf k}\sigma}  {\hat b}^{\;}_{{\bf k}{ \sigma}}
+\frac{B_{\bf k}}{2} 
\left({\hat b}^{\dagger}_{{\bf k}\sigma}  {\hat b}^{\dagger}_{-{\bf k}{\bar \sigma}}+{\rm H.c.}\right) \Big], 
\label{H3}
\end{eqnarray}
with $A_{{\bf k}{ \sigma}}=\left(\mu + s^2 \epsilon_{\bf k} - h_\sigma \right)$
and $B_{\bf k}=s^2\epsilon_{\bf k}$,
where $E_0\!=\!N(\mu\!-\!D) (s^2\!-\!1)$ is the bare ground-state energy, $N$ is the number of sites, $\sigma\!=\!\{\uparrow, \downarrow\}$, 
$h_\sigma\!=\!\pm h$, $\bar{\sigma}\!=\!-\sigma$, ${\hat b}^{(\dagger)}_{{\bf k}\sigma}$ are the Fourier transformed bosonic operators, and $\epsilon_{\bf k}\!=\!2 \sum_{\nu} J_{\nu} \cos{k_{\nu}}$. The anomalous terms indicate that bosons with opposite $S^z$ are created 
and annihilated in the ground state. These are the quantum fluctuations that lead to renormalization of the quasiparticle dispersion relation. The Hamiltonian (\ref{H3}) is diagonalized by the Bogolyubov transformation
\begin{eqnarray}
{\hat b}^{\;}_{{\bf k}\sigma} = u_{\bf k} {\beta}^{\;}_{{\bf k}\sigma}  + v_{\bf k} {\beta}^{\dagger}_{-{\bf k}{\bar \sigma}}\, ,
\label{uv}
\end{eqnarray}
 where
$u_{\bf k} v_{\bf k}= B_{\bf k}/2\omega^0_{\bf k}$,
$u^2_{\bf k}+v^2_{\bf k}=\left(\mu+s^2 \epsilon_{\bf k}\right)/\omega^0_{\bf k}$, and $\omega^0_{\bf k}=\sqrt{\mu^2+2 \mu s^2 \epsilon_{\bf k}}$. The resultant diagonal form of ${\hat {\cal H}}$ is
\begin{equation}
{\hat {\cal H}} = \widetilde{E}_0 + 
\sum_{{\bf k}} \Big[\left(\omega^0_{\bf k}-h\right)  
\beta^{\dagger}_{{\bf k}\uparrow} \beta^{\;}_{{\bf k}\uparrow} + 
\left(\omega^0_{\bf k}+h\right)  
\beta^{\dagger}_{{\bf k}\downarrow} \beta^{\;}_{{\bf k}\downarrow}\Big]\, . 
\label{Heff}
\end{equation}
Thus, the low-energy spectrum for $h\!<\!h_{c1}$ is 
$\widetilde\omega^<_{\bf k}\!\equiv\!\omega^0_{\bf k}\!-\!h$.
The band $\widetilde\omega^<_{\bf k}$ has a minimum at the antiferromagnetic wave-vector ${\bf Q}\!=\!(\pi,\pi,\pi)$ with the gap 
$\Delta^<\!=\!\omega^0_{\bf Q}-h$, whose vanishing point defines the critical field $h_{c1}\!=\!g \mu_BH_{c1}\!=\!\omega^0_{\bf Q}$. 
The ground state energy is also affected by quantum fluctuations: 
\begin{equation}
\widetilde{E}_0 = E_0 + \sum_{{\bf k}}
\left( \omega^0_{\bf k} - \mu - s^2 \epsilon_{\bf k}\right).
\end{equation}
The saddle point conditions, $ {\partial \widetilde{E}_0} / {\partial s }\!=\!{\partial \widetilde{E}_0} / {\partial \mu }\!=\!0$, lead to 
the self-consistent equations for the parameters $s$ and $\mu$, 
\begin{eqnarray}
s^2 = 2 - \frac{1}{N} \sum_{{\bf k}} \frac{\mu + s^2 \epsilon_{\bf k}}{\omega^0_{\bf k} },
\;\;\;
D = \mu + \frac{\mu}{N} \sum_{{\bf k}} \frac{\epsilon_{\bf k}}{\omega^0_{\bf k} }\, .
\label{smu}
\end{eqnarray}
Using the Hamiltonian parameters for DTN given in  Ref.~\cite{Zvyagin07}, the resulting values are $s^2=0.92$ and $\mu=10.3$K.

\begin{figure}[t]
\includegraphics[angle=0,width=0.8\columnwidth]{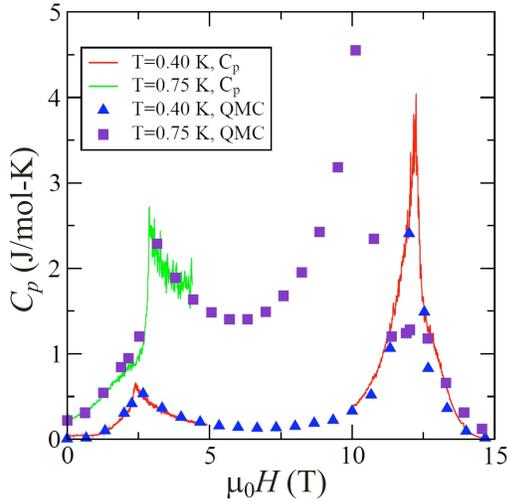}
\caption{Specific heat as a function of magnetic field for two  temperatures. Measurements and QMC simulations were also performed at the other temperatures showing nearly perfect agreement and the same characteristic behavior of $C_p$ vs $H$ (not included for clarity).}
\label{fig:fielddep}
\end{figure}

\begin{figure}[t]
\includegraphics[angle=0,width=0.8\columnwidth]{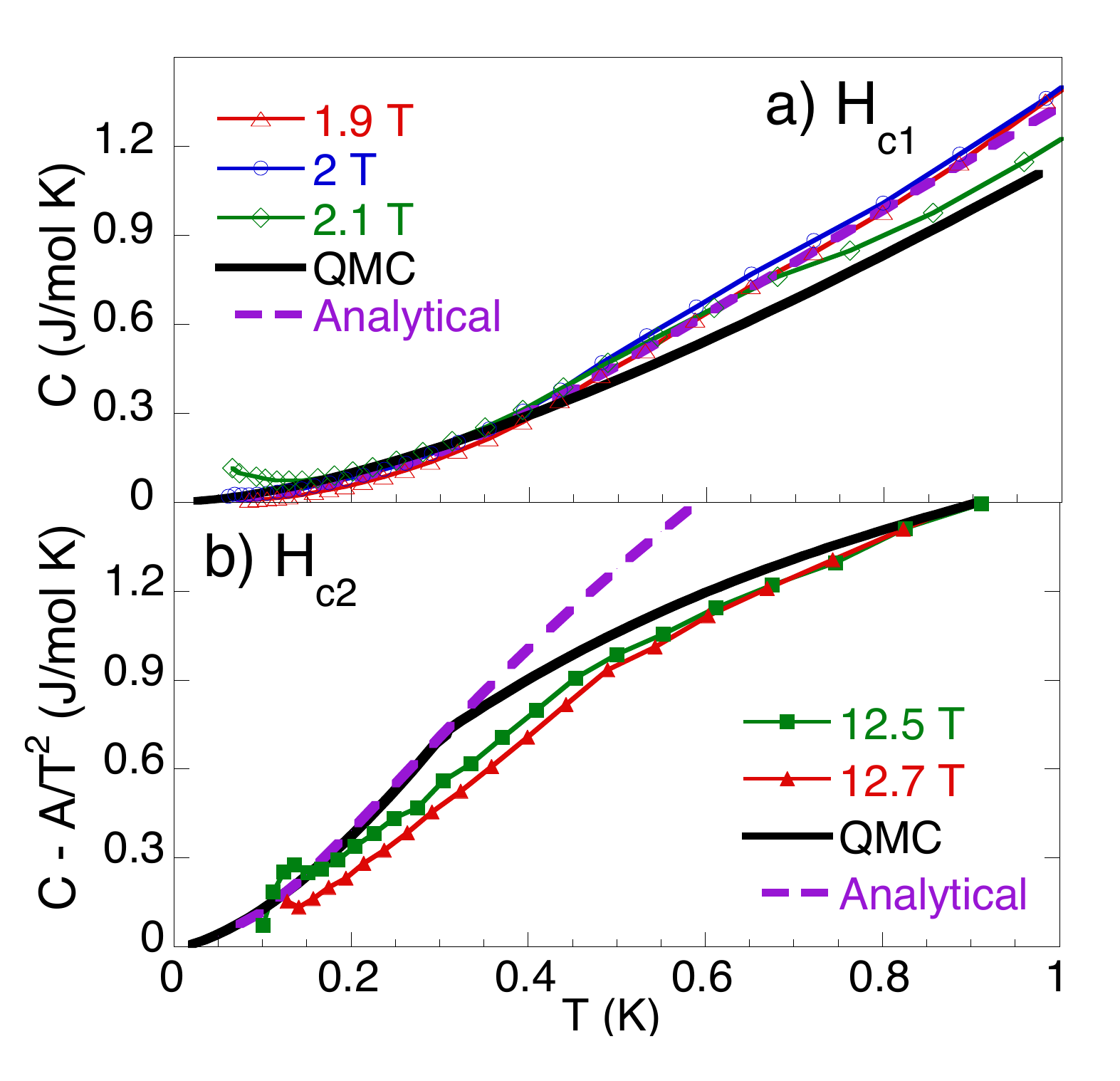}
\caption{Specific heat data 
as a function of  $T$ for  magnetic fields near 
\textbf{(a)} $H_{c1}$ and \textbf{(b)} $H_{c2}$. A Schottky anomaly tail, $A/T^2$ with $A = 0.018 J/mol K^3$, has been subtracted for fields near $H_{c2}$.  The full lines correspond to QMC simulations of Eq.(\ref{H1}) for $H=H_{c1(2)}$ and the parameters of Ref.~\cite{Zvyagin07}. The dashed lines are analytical calculations.}
\label{fig:lowTQCP}
\end{figure}

This low-energy theory  is  valid only for $H\!\leq\! H_{c1}$. For $H\!\geq\!H_{c2}$ spins are fully polarized and the
spectrum can be computed exactly. Since there are no quantum fluctuations for $H\!\geq\!H_{c2}$, the exact value of $h_{c2}$ is
$h_{c2}\!=\!g \mu_B H_{c2}\!=\!D-2 \epsilon_{\bf Q}$, while the unrenormalized excitation spectrum  is 
$\widetilde\omega^>_{\bf k}\!\equiv\!\epsilon_{\bf k}-\epsilon_{\bf Q}+h -h_{c2}$,
which also has a minimum at ${\bf Q}$ with the gap $\Delta^>\!=\!h-h_{c2}$.
Since only the excitations near ${\bf k}\!=\!{\bf Q}$ are important at low temperatures, we define the mass tensors for $H\!<\!H_{c1}$ 
and $H\!>\!H_{c2}$ as:
%from  $\widetilde\omega^<_{\bf k}$ and $\widetilde\omega^>_{\bf k}$, respectively:
\begin{equation}
\frac{1}{m^*_{\nu \nu}} = \frac{\partial^2 {\widetilde\omega^<_{\bf k}}} 
{{\partial k_{\nu}^2 }}\bigg|_{{\bf k=Q}}\, , \ \ \ \ \ \ \ \ 
\frac{1}{m_{\nu \nu}} = \frac{\partial^2 {\widetilde\omega^>_{\bf k}}} {{\partial k_{\nu}^2 }}\bigg|_{{\bf k=Q}}\, .
\label{ms}
\end{equation}
Then the mass renormalization factor is given by 
\begin{equation} 
\frac{m_{\nu\nu}}{m^*_{\nu\nu}} = s^2 \frac{\mu}{\omega^0_{\bf Q}}\approx
\frac{H_{c2}}{4H_{c1}}
\cdot\left(1+\sqrt{1+\frac{8H_{c1}^2}{H_{c2}^2}}\right)\, .
\label{mr}
\end{equation}
For the parameters of the Hamiltonian from  Ref.~\cite{Zvyagin07}, we obtain $m_{\nu\nu}/m^*_{\nu\nu} \simeq 3.2$. 
Such a large difference of masses must readily demonstrate itself in the strong asymmetry of the $C_v$ vs $H$ curves near $H_{c1}$ and $H_{c2}$ as  well as in the slopes of the specific heat dependence on $T$ 
at the critical fields where $C_v\!\propto\!\left(Tm\right)^{3/2}$.  These theoretical expectations are supported by the experimental 
$C_v(T,H)$ data shown in Figs.~\ref{fig:fielddep} and \ref{fig:lowTQCP}. 

The $C_p$ was measured in single crystals of DTN grown from aqueous solutions of thiourea and nickel chloride, with magnetic field applied 
along the crystalline $c$-axis. The experimental $C_p$ vs $H$ was obtained using an AC technique \cite{AC-Cp_Kohama}, while sweeping the magnetic 
field in a $^3$He fridge furbished with a 17~T superconducting magnet system at the National High Magnetic Field Laboratory (NHMFL) and the Los Alamos National 
Laboratory. We also used the standard thermal relaxation method to obtain $C_v$ vs $T$ with a dilution refrigerator in a 16~T Physical 
Properties Measurement System at Quantum Design, Inc. A strongly asymmetric $C_p$ vs $H$ is shown in Fig.~\ref{fig:fielddep} for fixed temperatures 
$T\!=\!0.75$K (green line) and $T\!=\!0.4$K (red line), alongside the results of the QMC simulations of the spin Hamiltonian in Eq.~(\ref{H1}) in a
$8 \times 8 \times 32 $ lattice (solid symbols). The agreement between the QMC results and the 
experimental data is very good. For the lowest temperature, the asymmetry of the peaks is $C_{p2}/C_{p1}\!\simeq\!6$, close to $(m/m^*)^{3/2}\!\approx\!5.7$ expected from the theory above.

Our $C_v$ vs $T$ experimental data close to $H_{c1}$ (top panel) and $H_{c2}$ (bottom panel) are displayed in Fig.~\ref{fig:lowTQCP} 
(lines with symbols). The results of the QMC simulations of Eq.\eqref{H1} at $H_{c1}$ and $H_{c2}$ are shown by the solid lines. The dashed lines correspond to the analytical calculation of $C_v(T)$ for a dilute gas of hard-core bosons  [Eq.~(\ref{Heff})] that have the spectra given by $\widetilde\omega^<_{\bf k}$ and $\widetilde\omega^>_{\bf k}$ at $H_{c1}$ and $H_{c2}$. The on-site boson-boson repulsion is taken into account 
at the mean-field level by summing the ladder diagrams (see Ref.~\cite{Nikuni00}). Since this approach is only valid for low density of bosons, it agrees closely with the QMC results at low $T$, but deviates from them at higher temperatures. The very good agreement between the theoretical results and the experimental curves  at $H_{c1}$ and $H_{c2}$ confirms {\it quantitatively} the expected mass renormalization for $H\!\leq\!H_{c1}$. 
%----------------------------------------------------------------------
%----------------------------------------------------------------------

The thermal conductivity was measured in DTN single crystals using the standard uniaxial heat flow method, where the temperature difference was 
produced by a heater attached to one end of the sample and monitored with a matched pair of RuO$_2$ thermometers. The heat flow 
and the magnetic field 
were
parallel to the $c$-axis. Similar observations were  
reported in Ref.~\onlinecite{Sun09} although their data at base temperature (380 mK)  do not agree with ours, measured down to 300 mK. 
%----------------------------------------------------------------------
%----------------------------------------------------------------------

\begin{figure}[b]
%\vspace{-0.5cm}
\includegraphics[angle=0,width=1.0\columnwidth]{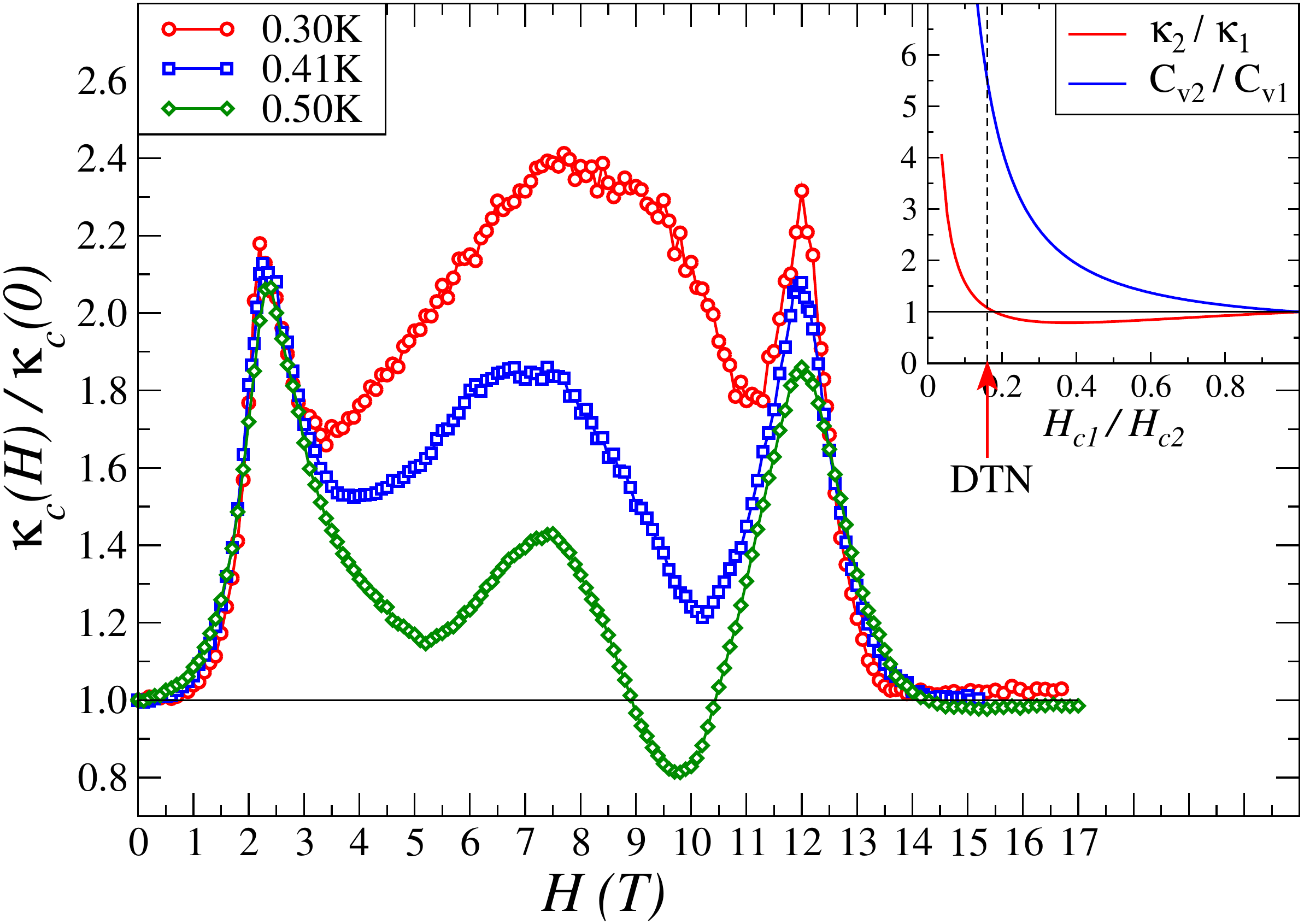}
\caption{Thermal conductivity of DTN along the \textsl{c}-axis as a function of magnetic field for several low-$T$ values. Inset: 
Theoretical prediction for the peak ratios $\kappa_2/\kappa_1$ and $C_{v2}/C_{v1}$ vs $H_{c1}/H_{c2}$.}
%\vspace{-0.7cm}
\label{fig:themalcond}
\end{figure}

The lighter mass of bosons for $H\!\leq\!H_{c1}$ implies not only large asymmetry between the peaks in the specific heat 
field dependence, but also similar asymmetries in a number of other properties of DTN  \cite{PaduanFilho04,Zvyagin07,Chiatti09,Zherlitsyn09,Zapf08}, 
all exhibiting a much stronger anomaly at $H_{c2}$ than at $H_{c1}$. In contrast, the  low-temperature thermal conductivity {\it does not} show any 
substantial asymmetry between the $H_{c1}$ and $H_{c2}$ data. Fig.~\ref{fig:themalcond} shows the field dependence of the thermal conductivity, 
$\kappa$, normalized to the $H\!=\!0$ value, $\kappa(0)$, for several low values of $T$. Since the $H\!=\!0$ magnetic excitation spectrum has a gap of about 
$3$K \cite{Zvyagin07}, only  phonons contribute to $\kappa(0)$ at low temperatures. The behavior of $\kappa$ changes qualitatively in the 
field because the gap  is closed between $H_{c1}$ and $H_{c2}$. The  low-temperature magnetic excitations provide a substantial contribution to 
the thermal conductivity as is clear from $\kappa(H)/\kappa(0)$ being $>\!1$ in Fig.~\ref{fig:themalcond}. Here we focus on the low-temperature 
behavior of $\kappa$ at the critical points $H_{c1}$ and $H_{c2}$. A detailed analysis of the other aspects of $\kappa$ will be 
provided elsewhere \cite{future1}.  
%----------------------------------------------------------------------
%----------------------------------------------------------------------

At low enough temperatures, scattering of bosons on each other should diminish because their concentration will be small. Consequently, the leading scattering in this regime should be due to defects. In the second Born approximation, the 
disorder-averaged inverse mean-free path of an excitation of mass $m$ due to scattering on point-like impurities is \cite{mahan1990}
\begin{eqnarray}
\label{1tau1}
\ell^{-1}=\frac{n_i}{2\pi}|V|^2 m^2 \, ,
\end{eqnarray} 
where $n_i$ is the impurity concentration and $V$ is the effective impurity potential.  When the excitation gap vanishes at $H_{c1}$ or $H_{c2}$, thermal conductivity at low-$T$ can 
be written as 
\begin{eqnarray}
\label{kappa0}
\kappa\propto\frac{\ell}{m}\int_0^{\sqrt{mT}} k^3 dk \propto
m T^2\cdot\ell
\propto\frac{T^2}{n_i m |V|^2}
\, .
\end{eqnarray} 
The theoretical temperature dependence, $\kappa\!\propto\!T^2$, is  in a good agreement with the measured low-$T$ thermal conductivity.
However, this dependence does not shed any light on the lack of asymmetry between the peaks. 
%as any type of disorder will yield that answer. 
Since vacancies or substitutional impurities are expected to be rare in clean
enough systems like DTN, we can assume that random lattice distortions are the most common source of disorder.
Because $D$ is the largest  parameter in Eq.\eqref{H1},  the most significant effect of these distortions is a 
real space modulation of $D$:
\begin{eqnarray}
\label{Himp}
 {\cal H}_{\rm imp}^D=\delta D(S^z_{\bf i})^2\Rightarrow
\delta D\sum_{\sigma,{\bf k,k'}} 
e^{i{\bf R}_{\bf i} ({\bf k-k'})}\, b_{{\bf k}\sigma}^\dag
 b^{\phantom \dag}_{{\bf k'}\sigma} \, ,
\end{eqnarray} 
where ${\bf i}$ is the impurity site and we have used the mapping (\ref{sop}). 

This is where the renormalization due to quantum fluctuations becomes crucial again. For $H\!=\!H_{c2}$ the impurity scattering in (\ref{Himp}) is not renormalized  due to the absence of quantum fluctuations and $V_2\!\equiv\!\delta D$. On the other hand, for  $H\!=\!H_{c1}$ the scattering {\it is} affected by the quantum fluctuations. Since the dressed bosonic excitations are related to the bare ones through Eq.~(\ref{uv}), this transforms impurity scattering (\ref{Himp}) into
\begin{eqnarray}
\label{Himp1}
{\cal H}_{\rm imp}^D=\delta D\sum_{\sigma,{\bf k,k'}} 
e^{i{\bf R}_\ell ({\bf k-k'})}
\left(u_{\bf k}u_{\bf k'}+v_{\bf k}v_{\bf k'}\right)
\beta^\dag_{{\bf k}\sigma}\beta_{{\bf k'}\sigma} \, . 
\end{eqnarray} 
Thus, the impurity potential  at $H_{c1}$ is 
$V_1\!=\!\delta D\left(u_{\bf Q}^2\!+\!v_{\bf Q}^2\right)$, which will modify the mean-free path in (\ref{1tau1}). After some algebra  utilizing Eqs.~(\ref{kappa0}), (\ref{uv}) and (\ref{mr}), we finally obtain
\begin{eqnarray}
\label{kappa_r1}
\frac{\kappa_2}{\kappa_1}=
\frac{m\cdot\ell_2}{m^*\cdot\ell_1}=
\left(\frac{m}{m^*}\right)
\cdot\frac{1}{4s^4}\cdot\left(1+s^4\left(\frac{m^*}{m}\right)^2\right)^2
\, .
\end{eqnarray} 
This expression contains a large prefactor  $(m/m^*)$ coming from the renormalization of the density of states and velocity in (\ref{kappa0}),  and would formally imply a larger peak at $H_{c2}$, similar to the specific heat and other quantities. However, this effect is partially compensated 
by the numerical factor $\approx\!1/4\!+\!O((m^*/m)^2)$, which comes from the renormalization of the mean-free path. 
By using the DTN parameters, we obtain 
$\kappa_2/\kappa_1\!\approx\!1.1$ in an excellent agreement with the data in Fig.~\ref{fig:themalcond}.
Thus, the mass renormalization effect in thermal conductivity is compensated by a similar renormalization effect in the impurity scattering. 
To show that this is not a mere coincidence, we provide our prediction for the $H_{c1}/H_{c2}$ dependence 
of the peak ratios in thermal conductivity $\kappa_2/\kappa_1$ and specific heat $C_{v2}/C_{v1}$ on $H_{c1}/H_{c2}$ (see inset in Fig.~\ref{fig:themalcond}). Here we used the relation between the mass ratio and $H_{c1}/H_{c2}$ given by Eq.~(\ref{mr}).
The vertical line corresponds to the DTN value of $H_{c1}/H_{c2}\!\approx\!0.17$. 
%The full dependence on $H_{c1}/H_{c2}$ is important for extending our results to other materials. 
It is remarkable that  
$\kappa_2/\kappa_1$ and $C_{v2}/C_{v1}$ behave in very different ways. In particular, 
$\kappa_2/\kappa_1\!\approx\!1$ while the peaks in $C_v$ are very asymmetric for $0.1\!\lesssim\! H_{c1}/H_{c2}\!\leq\!1$. With this insight, we also suggest an experimental verification of our theory by conducting the heat conductivity measurement in DTN under pressure. A modest decrease of $H_{c1}$ by 1T should lead to an increase in $\kappa_2/\kappa_1$ by a factor of 2.

The leading impurity scattering (\ref{Himp}) and, consequently, the resulting expression for the ratio $\kappa_2/\kappa_1$ in 
(\ref{kappa_r1}) will remain valid for the other BEC magnets even though they may not be dominated by the single-ion anisotropy term. For instance, in the dimer-based systems \cite{Jaime04}, the disorder in the leading intra-dimer coupling translates into the local modulation of the chemical potential  which is equivalent to our Eq.~(\ref{Himp}).  Thus, our Eq.~(\ref{kappa_r1}) can be verified  in other BEC compounds. 

In conclusion, by using the example of DTN, we connected the  asymmetry in the physical properties of BEC magnets with 
the mass renormalization of the elementary excitations due to quantum fluctuations of the paramagnetic state. We also resolved the enigmatic 
absence of this asymmetry in the low-$T$ thermal conductivity by identifying the leading scattering mechanism and by demonstrating that the 
renormalization of the latter compensates the mass renormalization effect.

This work was supported by the NSF, the State of Florida, the US DOE  
under grant DE-FG02-04ER46174 (A. L. C.) and by the DFG, SFB 608 (A.S. and J.M.).

%\bibliography{DTN_a}

\begin{thebibliography}{20}
\bibitem[{not()}]{note}
\bibinfo{note}{The number of bosons or uniform magnetization is never strictly
  conserved in real magnets due to spin-orbit interaction.}
\end{thebibliography}

\end{document}